\documentclass{pasj00}
\draft

\begin{document}
\SetRunningHead{H. Matsuhara et al.}{Optical Identification of 15 Micron Sources)}
\Received{          }
\Accepted{          }

\title{Optical Identification of 15 Micron Sources in the AKARI Performance Verification Field toward the North Ecliptic Pole}

%
%
\author{%
   Hideo \textsc{Matsuhara},\altaffilmark{1}
    \thanks{Further information contact Hideo Matsuhara (maruma@ir.isas.jaxa.jp)}
   Takehiko \textsc{Wada},\altaffilmark{1}
   Chris P. \textsc{Pearson},\altaffilmark{1,10} 
   Shinki \textsc{Oyabu},\altaffilmark{1} \\
   Myungshin \textsc{Im},\altaffilmark{2}
   Koji \textsc{Imai},\altaffilmark{1} 
   Toshinobu \textsc{Takagi},\altaffilmark{1}
   Eugene \textsc{Kang},\altaffilmark{2} \\
   Narae \textsc{Hwang},\altaffilmark{2} 
   Woong-Seob \textsc{Jeong}, \altaffilmark{1}
   Hyung Mok \textsc{Lee},\altaffilmark{2} 
   Myung Gyoon \textsc{Lee},\altaffilmark{2} \\ 
   Soojong \textsc{Pak},\altaffilmark{3} 
   Stephen \textsc{Serjeant},\altaffilmark{4}
   Takao \textsc{Nakagawa},\altaffilmark{1} 
   Hitoshi \textsc{Hanami},\altaffilmark{5} \\
   Hanae \textsc{Inami},\altaffilmark{1} 
   Takashi \textsc{Onaka},\altaffilmark{6} 
   Naofumi \textsc{Fujishiro},\altaffilmark{1}\thanks{Present address : Cybernet systems Co. Ltd., Bunkyo-ku, Tokyo 112-0012, Japan}
   Daisuke \textsc{Ishihara},\altaffilmark{6} \\ 
   Yoshifusa \textsc{Ita},\altaffilmark{1} 
   Hirokazu \textsc{Kataza},\altaffilmark{1}  
   Woojung \textsc{Kim},\altaffilmark{1}
   Toshio \textsc{Matsumoto},\altaffilmark{1} \\
   Hiroshi \textsc{Murakami},\altaffilmark{1} 
   Youichi \textsc{Ohyama},\altaffilmark{1} 
   Itsuki \textsc{Sakon},\altaffilmark{6} 
   Toshihiko \textsc{Tanab\'{e}},\altaffilmark{7} \\
   Kazunori \textsc{Uemizu},\altaffilmark{1} 
   Munetaka \textsc{Ueno},\altaffilmark{8} 
   and
   Hidenori \textsc{Watarai}\altaffilmark{9} 
   }
 \altaffiltext{1}{Institute of Space and Astronautical Science, Japan Aerospace Exploration Agency, \\
   Sagamihara, Kanagawa 229 8510 } 
 \altaffiltext{2}{Department of Physics \& Astronomy, FPRD, Seoul National University, \\
  Shillim-Dong, Kwanak-Gu, Seoul 151-742, Korea}
 \altaffiltext{3}{Kyung Hee University, 1 Seocheon-dong, Giheung-gu, Yongin-si
    Gyeonggi-do 446-701, Korea} 
 \altaffiltext{4}{Astrophysics Group, Department of Physics, The Open University, Milton Keynes, MK7 6AA, UK}
 \altaffiltext{5}{Iwate University, 3-18-8 Ueda, Morioka, 020-8550 } 
 \altaffiltext{6}{Department of Astronomy, School of Science, University of Tokyo, Bunkyo-ku, Tokyo 113-0033 }
 \altaffiltext{7}{Institute of Astronomy, University of Tokyo, Mitaka,Tokyo 181-0015}
 \altaffiltext{8}{Department of Earth Science and Astronomy, Graduate School of Arts and Sciences,\\
   The University of Tokyo, Meguro-ku,Tokyo 153-8902}
 \altaffiltext{9}{Office of Space Applications, Japan Aerospace Exploration Agency, Tsukuba, Ibaraki 305-8505 }
 \altaffiltext{10}{ISO Data Centre, ESA, Villafranca del Castillo, Madrid, Spain.}


\maketitle

\begin{abstract}
We present the results of optical identifications for 257 mid-infrared sources
detected with a deep 15~$\mu$m survey over approximately 80~arcmin$^2$ area in
the AKARI performance verification field near the North Ecliptic Pole. The 15~$\mu$m fluxes 
of the sources range from 1~mJy down to 40~$\mu$Jy, approximately a half of which are
below 100~$\mu$Jy. Optical counterparts were searched for within a 2-3~arcsec radius
in both the $BVRi'z'$ catalog generated by using the deep Subaru/Suprime-cam field 
which covers one-third of the performance verification field, and the $g'r'i'z'$ catalog
based on observations made with MegaCam at CFHT.
We found $B-R$ and $R-z'$ colours of sources with successful optical identifications are 
systematically redder than that of the entire optical sample in the same field. Moreover, 
approximately 40\% of the 15~$\mu$m sources show colours $R-L15>5$, which cannot be explained by 
the  spectral energy distribution~(SED) of normal quiescent spiral galaxies, but are consistent 
with SEDs of redshifted ($z>1$) starburst or ultraluminous infrared galaxies. This result 
indicates that the fraction of the ultraluminous infrared galaxies in our faint 15~$\mu$m sample 
is much larger than that in our  brighter 15~$\mu$m sources, which is consistent with the 
evolving mid-infrared luminosity function derived by recent studies 
based on the Spitzer 24~$\mu$m deep surveys.  Based on an SED fitting technique, the nature of the faint 15~$\mu$m sources is further discussed 
for a selected number of sources with available $K_{s}$-band data. 

\end{abstract}

\section{Introduction}\label{sec:introduction}

Finding new populations of faint high-redshift galaxies is an important step to unveil the Cosmic
star formation history of the Universe. Deep mid-infrared (especially at 15~$\mu$m) and far-infrared (at 90, 170~$\mu$m)
surveys with the Infrared Space Observatory (ISO, \cite{GeCe00}, and references therein) discovered a distinct
population with faint optical fluxes and large infrared luminosities ($L_{\rm bol} \geq 10^{11} L_{\odot}$,
\cite{elb02}) at $z \geq 0.5$. The 15~$\mu$m source counts obtained with the ISO surveys implied strong
evolution in the galaxy population (\cite{GeCe00}, \cite{aus99}, \cite{ser01}, \cite{gru02}, \cite{oli02}).
Not only the source count models for mid-IR counts but also the X-ray data~\citep{mane04} and subsequent 
spectroscopic studies~\citep{mrr04} confirmed that the bulk of the ISO sources were starburst galaxies not the AGNs.

The {\it Spitzer Space Telescope}~\citep{wer04} also offers excellent  sensitivities
in the four IRAC wavebands between 3.6 and 8.0~$\mu$m~\citep{faz04} and at MIPS 24~$\mu$m~\citep{rie04}, 
and has revealed that the key epoch to understand the formation and evolution of massive 
($M \geq 10^{11}M_{\odot}$) galaxies is at $z=1-2$: the star formation activity of massive galaxies 
at $z \sim 2$ is hundred times larger than that at present ~(\cite{cha06}, \cite{capu06}). These 
ultra-luminous ($L\geq 10^{12}L_{\odot}$) infrared galaxies~(ULIRGs) may be the progenitors of present-day
 massive galaxies. In order to reveal their nature, we require complete, uniform samples of massive star
 forming galaxies at $z=1-2$.
Although the {\it Spitzer} 24~$\mu$m deep surveys can probe galaxies
at this redshift range, deep surveys at shorter mid-infrared wavelengths
(10-20~$\mu$m, in the gap between the IRAC 8~$\mu$m band and the MIPS 
24~$\mu$m band) are very important to unveil the nature of them, since
the strong 6.2 \& 7.7~$\mu$m PAH features prominent in the mid infrared spectra 
of star-forming galaxies can enter the MIPS 24~$\mu$m passband only at $z > 2$.
{\it Spitzer}/IRS~\citep{hou04} has imaging capability at 16~$\mu$m via its peak-up camera, and with this camera deep 16~$\mu$m 
surveys over 150~arcmin$^2$ in each of GOODS-North and GOODS-South fields have been carried out to depths of 50-85~$\mu$Jy~(3$\sigma$)~(\cite{tep07},
\cite{teplitz05}). However, the areal coverage is not yet comparable to the surveys at 24~$\mu$m due to the relatively small field of view (1.2 arcmin$^2$) of the IRS peak-up camera. 

The AKARI satellite, launched on February 21, 2006 (UT), has the capability for deep mid-infrared imaging
 in the {\it Spitzer} wavelength desert between  8~$\mu$m and 24~$\mu$m through one of its focal plane instruments, 
the InfraRed Camera (IRC, \cite{ona07}). The IRC incorporates three infrared cameras covering nine bands between 
2 \& 24~$\mu$m suitable for deep cosmological surveys. Note that due to the nature of the orbit of AKARI 
(Sun-synchronous), the visibility of any point on the sky is a strong function of ecliptic latitude and thus deep surveys 
are only possible at the ecliptic poles. The AKARI North Ecliptic Pole (NEP) survey is a major legacy of the AKARI mission 
consisting of a deep 0.4 square degree and shallow 6.2 square degree survey in all 9 IRC bands \citep{maruma06}. 
The AKARI NEP surveys, especially at 11, 15, and 18~$\mu$m, are well matched to the  {\it Spitzer}  24~$\mu$m 
surveys (e.g., \cite{papovich04}), sampling similar cosmological volumes and are more sensitive to high redshift star-formation 
activity than the shorter wavelength  {\it Spitzer}/IRAC  8~$\mu$m band.

In this paper we describe a selection of initial results from the deep extragalactic survey around the NEP region, focusing on the optical nature of the faint 15~$\mu$m sources detected in the performance verification phase of the AKARI mission. The 15~$\mu$m sample is especially unique since more than 100 sources are fainter than 100~$\mu$Jy, a limit below that obtained with ISO. In section~\ref{sec:data}, we briefly describe the AKARI data as well as the optical data obtained with Subaru/Suprime-cam and CFHT/Megacam, and the results of identification of the optical counterparts. In section~\ref{sec:discussion}, we discuss the nature of the 15~$\mu$m sample based on the optical -- mid infrared colours. Section~\ref{sec:summary} gives the summary of the paper. Throughout the paper we use the AB maginitude system, unless otherwise explicitly noted : 20 AB mag corresponds to 36$\mu$Jy. We adopt a cosmology of $\Omega_{m}=0.3$, $\Omega_{\Lambda}=0.7$, and $H_{0}=70$km sec$^{-1}$ Mpc$^{-1}$.

\section{The data and the Results of Identification}\label{sec:data}

\subsection{{\it AKARI}/IRC mid-infrared data}

The NEP Deep survey is centred on a circle at R.A. = 17$^h$55$^m$24$^s$, 
dec = +66$^{\circ}$37$^{\prime}$32$^{\prime\prime}$. However, during the performance 
verification phase (13th April -- 8th May 2006) of the AKARI mission  a pilot survey 
of the NEP over a single field of view of the IRC (approximately 10$^{\prime} \times $10$^{\prime}$, (hereafter referred to as the performance verification field) at R.A. 
= 17$^h$57.$^m$3, dec = +66$^{\circ}$54.$^{\prime}$3), ten pointings 
deep in the IRC L15 band (i.e. in the 15$\mu$m band) was carried out. 

 Here we briefly report the observations and the data reduction: full details are described in \citet{wada07a}. The data for this work were taken using the IRC05 Astronmoical Observation Template (AOT) mode which is optimized for deep survey observations. Note that the IRC05 AOT minimizes overheads by using the minimum number of resets, no filter change and moreover no dithering operation during the 10 minute integration time for one pointing. Therefore, the dithering is performed among the individual pointing
observations to remove the effects of dead/hot pixels and cosmic rays, etc. The total net exposure time 
of the ten pointing observations was 4417 seconds.
 The data from individual frames for each pointing were reduced using the standard  IRC data reduction pipeline version 060801 \citep{ita07} within the IRAF environment \footnote{IRAF is distributed by the National Optical Astronomy Observatory, which is operated by the Association of
Universities for Research in Astronomy, Inc., under cooperative
agreement with the National Science Foundation.}. The IRC pipeline splits pointings into individual frames and corrects for instrument characteristics by masking of anomalous/dead pixels, 
and then applies dark subtraction, linearity correction, saturation, distortion correction and flat fielding, etc. 
Astrometric data (world coordinate system, WCS) is applied by matching bright stars within the data with 2MASS counterparts. 
Identification of bright point sources in the deep optical image~(\cite{wada07b}, see section~\ref{sec:optical}) by eye suggests 
that the positional accuracy of the WCS is much better than two pixels (5~arcsec). The resulting stacked images for the individual 
pointing observations are then co-added by identifying the relatively bright sources in each processed image of one pointing. 
Finally the edge of the image, where the signal-to-noise ratio is  worse, was trimmed, resulting in a  final image size of 77.29~arcmin$^{2}$. 
We used SExtracter~\citep{BeAr96} for the source detection, and source fluxes were evaluated by aperture photometry with a radius of 1.3~arcsec.
 In total 257 sources were extracted between $L15$=16 and 20~magnitudes, of which 110 sources were fainter than 19~mag.

At present, the depth of the final L15 image is approximately 42~$\mu$Jy in 3$\sigma$, probably due to the fact that the source extraction technique is not yet optimized. The completeness analysis result implies that the number of sources below 150~$\mu$Jy is substantially underestimated, but once detected, the source detection is reliable down to $L15$=20~mag~(or 36 $\mu$Jy; \cite{wada07a}).  This is justified since 60--80\% of the sources fainter than $L15$=18~mag can be identified in the optical images~(see section~\ref{sec:subaru_id}).  

\subsection{Optical Data}\label{sec:optical}

 Approximately one third of the performance verification field is covered by a single Subaru Suprime-cam 
field of view (916 arcmin$^2$) to $B$=28.4, $V$=27, $R$=27.4, $i^{\prime}$=27, $z^{\prime}$=26.2 
(3$\sigma$, AB magnitude, \cite{wada07b}, in preparation). 
The observations were carried out in June and September 2003 with a typical seeing of 1.0~arcsec.

 The remaining two thirds of the performance verification field is covered by $g^{'}$, $r^{'}$, $i^{'}$, 
$z^{'}$-band images taken with the MegaCam instrument on the Canada-France-Hawaii Telescope (CFHT).
 The MegaCam observation was carried out in August/September 2004  to support the AKARI NEP survey. 
The observed field covers roughly  a $2^{\circ}$ by $1^{\circ}$ field of view centered on the NEP,  
therefore the whole performance verification field is covered
by the CFHT imaging data. The depth of the CFHT image is estimated to be $g^{'} \sim 26.4$ mag,
 $r^{'} \sim 25.9$ mag, $i^{'} \sim 25.3$ mag, and $z^{'} \sim 24.0$ mag
 at 3 $\sigma$ over an aperture with $1\farcs0$ diameter, but 
 effectively the source counts are complete only down to a 
 limit of about 1.5 magnitudes brighter than the above  numbers. The absolute astrometric accuracy 
of the CFHT mosaic data is measured to be about r.m.s $\sim$ $0\farcs4$.  More details on the CFHT 
images can be found in \citet{Hwa07}.

\subsection{Identification with Subaru/Suprime-cam source catalog}\label{sec:subaru_id}

The result of the identification within a  2~arcsec search radius is shown in Table~\ref{tab:id_sum_subaru}.
Due to the partial coverage of the Subaru image, only 105 sources are inside the Subaru image. 
Between $L15$=16 and 18~magnitudes, all the three L15 sources without optical matches
 are found to be a blend of several sources or a part of a large, bright galaxy. Blend
 of several sources in L15 causes the coordinate to be calculated as the mean of the 
 multiple sources, making it escape from the 2~arcsec matching radius. 
 Between $L15$=18 and 19~magnitudes, six out of seven L15 sources without optical
 matches are found to be a part of a bright optical counterpart, or located near the 
 edge, or a blend of multiple sources. The remaining source is found near the edge of
 the L15 image. It could be a genuine mid-infrared source with no optical counterpart,
 but it is located near the edge of the L15 image where the signal-to-noise ratio is low.
 Between $L15$=19 and 20~magnitudes, 12 souces out of 15 with no optical 
 counterpart are again, blend of multiple sources, inside optically bright galaxies, 
 or probably spurious objects near the edge of the L15 image. The remaining three L15 
 sources may be genuine optically faint sources. Overall, of 24 L15 sources without 
 optical matches, only three or four are found to be sources without optical counterparts
 which are worth to be investigated carefully by multi-colour images taken during the
 NEP survey program. The rest are L15 detection of multiple sources, sources within
 optically bright galaxies, or erroneous matches near the edges.

Figure~\ref{color_mag} shows their $R-L15$ colour with 15~$\mu$m magnitude. Except for the 
very blue sources which are found to be bright stars (point-like sources in the Subaru image), 
the $R-L15$ colour distribution does not change with 15~$\mu$m magnitude and has a median value 
of $R-L15$=4.6~mag.

\begin{figure}[htbp]
  \begin{center}
    \FigureFile(102mm,133mm){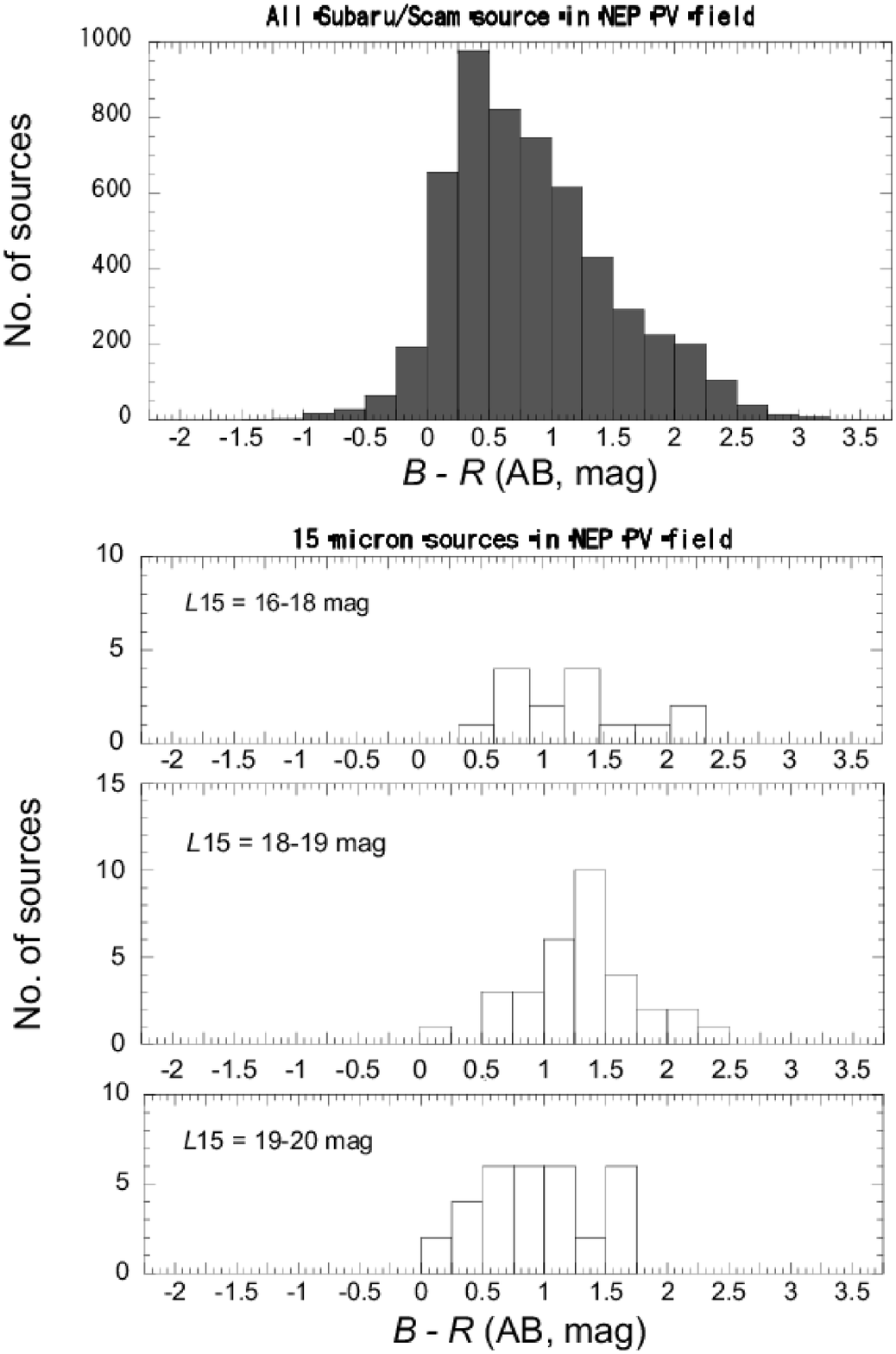}
  \end{center}
  \caption{
 The observed optical ($B-R$ ) color for all $z'$-band selected 
 Subaru/Suprime-cam source catalog in the field (top)
 and for the 15~$\mu$m sources with successful identification on 
 the Subaru/Suprime-cam source catalog(bottom).
 }\label{src_hist}
\end{figure}

\begin{figure}[htbp]
  \begin{center}
    \FigureFile(102mm,133mm){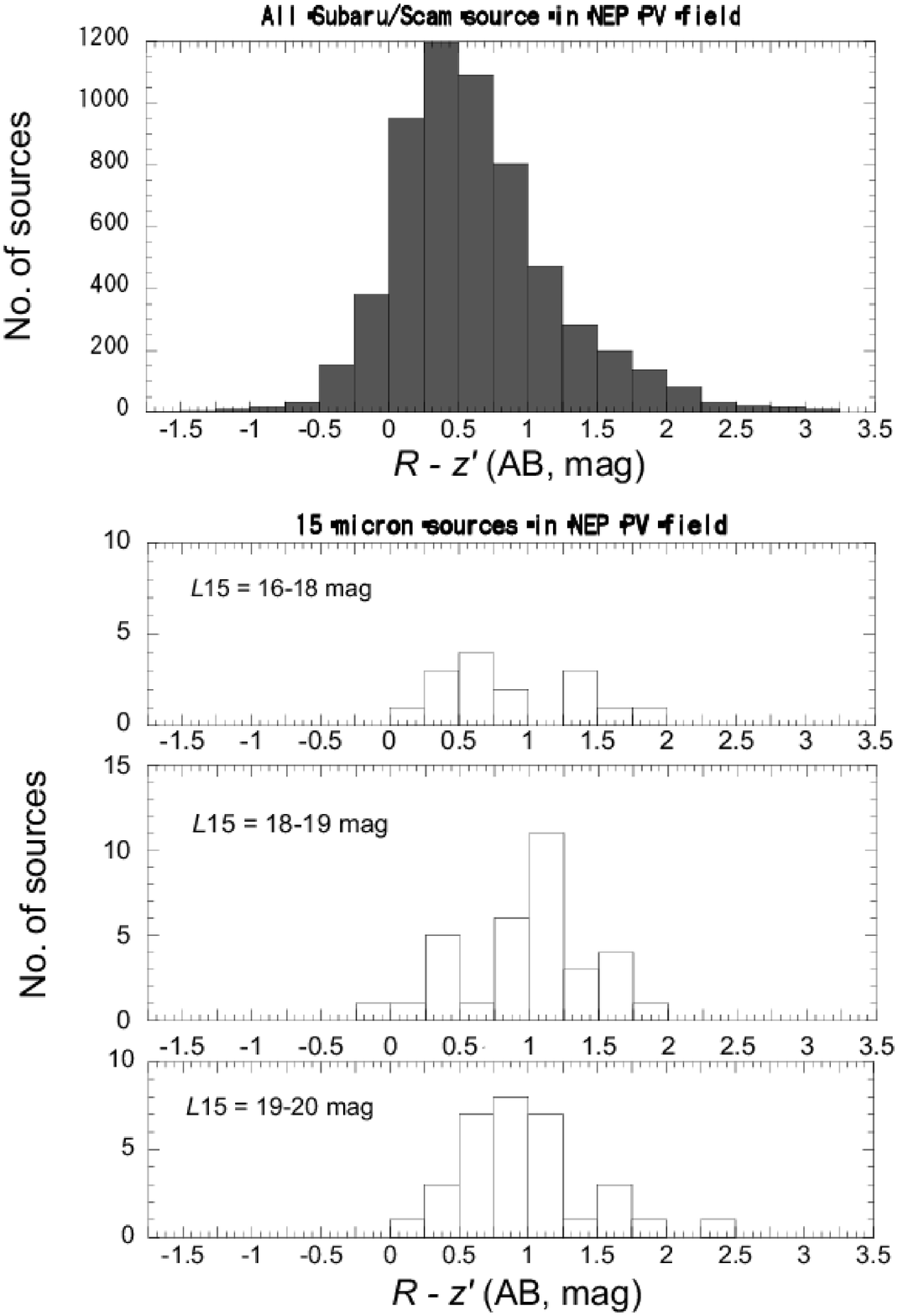}
  \end{center}
  \caption{
The observed optical ($R-z'$ ) color for all $z'$-band selected Subaru/Suprime-cam source catalog in the field (top)
and for the 15~$\mu$m sources with successful identification on the Subaru/Suprime-cam source catalog(bottom).
 }\label{src_hist2}
\end{figure}

As examples showing their optical colour characteristics, in Figure~\ref{src_hist},~\ref{src_hist2} we
show a comparison of the $B-R$, $R-z'$ colours between all the Subaru sources in the NEP performance verification field and the 15~$\mu$m sources. 
The optical colours of the 15~$\mu$m sources show a clear trend towards a redder colour : 0.3--0.4 mag redder in $B-R$, and 0.2-0.5 mag in $R-z'$. 
This indicates that the 15~$\mu$m sources sample either a relatively
high-redshift population or a population exhibiting dust reddening. Moreover, the fainter 15~$\mu$m sources at or
below 18~mag(AB) show redder $R-z'$ colour (approximately 0.3~mag) than the brigher sources. This may indicate that
fainter 15~$\mu$m sources are at relatively high redshift, since the shift of their Balmer/4000A spectral break in
between the $R$ and $z'$ bands will create redder  $R-z'$ colours.

\begin{table}
  \caption{Summary of Identification in the Subaru Image (2~arcsec search radius)}\label{tab:id_sum_subaru}
  \begin{center}
    \begin{tabular}{llll}
	\hline \hline
15~$\mu$m  & No. of total 15~$\mu$m & No. of sources inside  & No. of sources with       \\ 
magnitude  &  sources               & the Subaru image       & successful identification \\ \hline
16--18     &    44                  &  18                    &    15(0)\footnotemark[\#] \\ 
18--19     &    103                 &  40                    &    33(2)                  \\ 
19--20     &    110                 &  47                    &    32(7)                  \\ \hline
\multicolumn{4}{l}{\parbox{125mm}{\footnotesize \noindent
  \footnotemark[\#] All 18 sources have conterparts within 3~arcsec search radius. 
  \par \noindent
  Numbers in the parenthesis are those with two or more optical counterparts. 
}}
    \end{tabular}
  \end{center}  
\end{table}

\subsection{Identification with CFHT/Megacam source catalog}

 The result of the cross-identification with 2\arcsec~ search radius for optical sources in the
CFHT images is shown in Table~\ref{tab:id_sum_cfht}.

\begin{table}
  \caption{Summary of Identification in the CFHT Image}\label{tab:id_sum_cfht}
  \begin{center}
    \begin{tabular}{llll}
	\hline \hline
15~$\mu$m  & No. of sources inside & No. of sources in      & No. of sources in      \\ 
magnitude  & the CFHT image        & 3~arcsec search radius & 2~arcsec search radius \\ \hline
16--18     &  42                   & 39(13)                 & 27(3)                  \\ 
18--19     &  99                   & 61(10)                 & 50(5)                  \\ 
19--20     &  108                  & 67(7)                  & 49(4)                  \\ \hline
\multicolumn{3}{l}{\parbox{125mm}{\footnotesize \noindent 
 Numbers in the parenthesis are those with two or more optical counterparts.  
}}
    \end{tabular}
  \end{center}  
\end{table}

\begin{figure}[htbp]
  \begin{center}
    \FigureFile(146mm,120mm){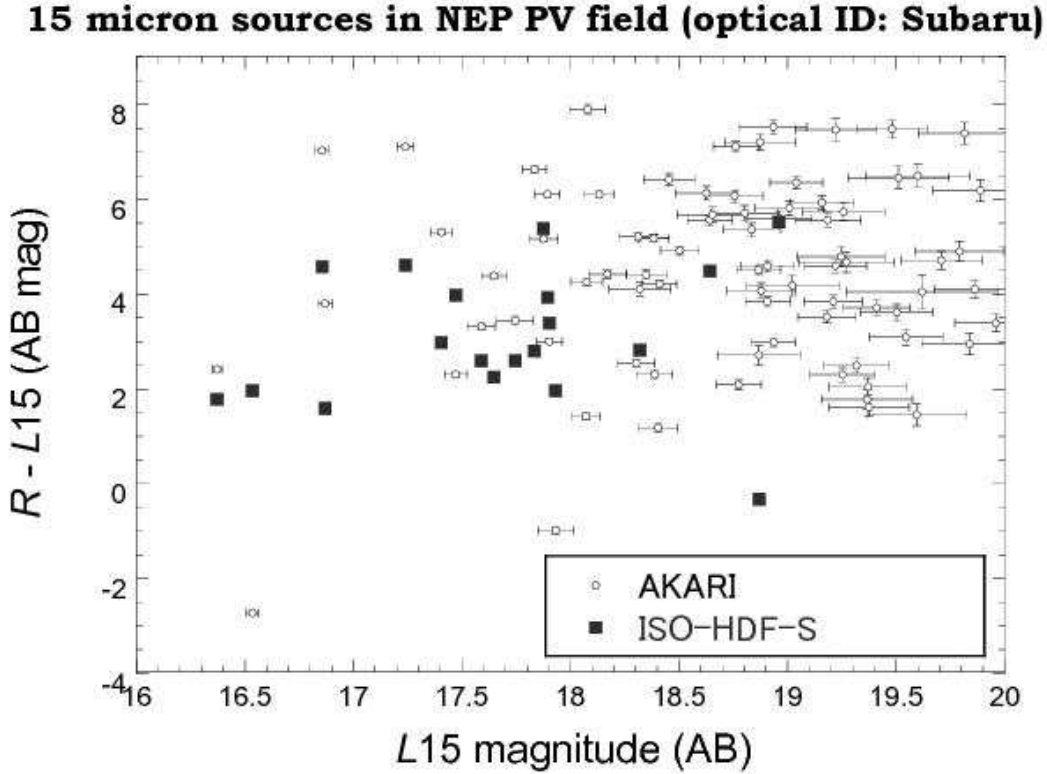}
  \end{center}
  \caption{
The observed optical -- mid-infrared color vs 15$\mu$m magnitude(AB) for all the 15~$\mu$m sources with 
successful identification on the Subaru/Suprime-cam source catalog. Black squares represent the HDF-S sources detected with the ISOCAM LW3 15~$\mu$m deep 
survey~(\cite{oli02},~\cite{man02}). 
 }\label{color_mag}
\end{figure}

  We find optical counterparts for 93\% of the bright L15 sources ($16 < L15 < 18$ mag),
 which is consistent with the matching result using the Subaru data.
 A large fraction of sources with a positional offset between 2\arcsec~ and 
 3\arcsec~ turn out to be multiple optical sources blended into a single 
 object in the L15 image. 
  For the fainter L15 sources, the matching probability drops to 
 $\sim$65\% with a 3\arcsec~ matching radius and down to 50\% when we use a
 2\arcsec~ matching radius. The above number is about 10-30 \% less than
 the matching probability of 60 -- 80 \% for L15 sources in the Subaru image.
 The discrepancy mainly comes from the fact that about 10\% of L15 sources have
 optical magnitudes around $R\sim 26$ (Figure~\ref{color_mag}) for which no $r^{\prime}$-band
counterparts are found in the CFHT image. Thus, although the CFHT image covers the entire performance verification field, we only use the sources in the Subaru image 
for the discussion of the nature of the faint 15~$\mu$m sources.

  One should note that the simple automatic identification presented here also
 suffers from the chance coincidence.
 Based on the number counts of $r^{\prime}$-band sources presented in \citet{Hwa07},
 the chance coincidence of source identification is 6\% and 15\% for 3\arcsec~ search radius for sources
 at $r^{\prime} < 22$ and  $r^{\prime} < 24$, respectively. This number drops to 3\% and 7\% for a 2\arcsec~ search radius. 
  Clearly, the 2\arcsec~ search radius suffers much less from chance
 coincidence. However, a 2\arcsec~ radius may be too small in the case for some 15~$\mu$m sources where 
 the mid-infrared flux originates from multiple sources.

\subsection{Identification with KPNO/Flamingos near-infrared source catalog}

The Subaru/Suprime-cam field around the NEP is also covered by the KPNO~2.1m/FLAMINGOS at $J$- and $K_{s}$-band to a depth of $J$=21.6 and $K_{s}$=19.9 (3$\sigma$) in Vega magnitudes (``Field-NE" in \cite{imai07}). 
Thus we also searched for near infrared counterparts of the 15$\mu$m sources with optical identifications in the Subaru/Suprime-cam images. 
However, the performance verification field is located at the edge of the survey field and thus the number of available images for stacking varies within the image, and also additionally suffers from image distortion from the FLAMINGOS camera. 
As a result, among the total 81 sources in Table~\ref{tab:id_sum_subaru}, only 27 and 28 sources are identified in the $J$- and $K_{s}$ bands respectively, with 23 sources being identified in both the  $J$- and $K_{s}$ bands. 
Since the near infrared data is incomplete, we only utilize the near infrared magnitudes for the investigation of several selected sources with relatively reliable identifications~(see section~\ref{sec:nature}).

\section{Discussion}\label{sec:discussion}

\subsection{Colours of the AKARI 15$\mu$m population}

\begin{figure}[htbp]
  \begin{center}
    \FigureFile(126mm,126mm){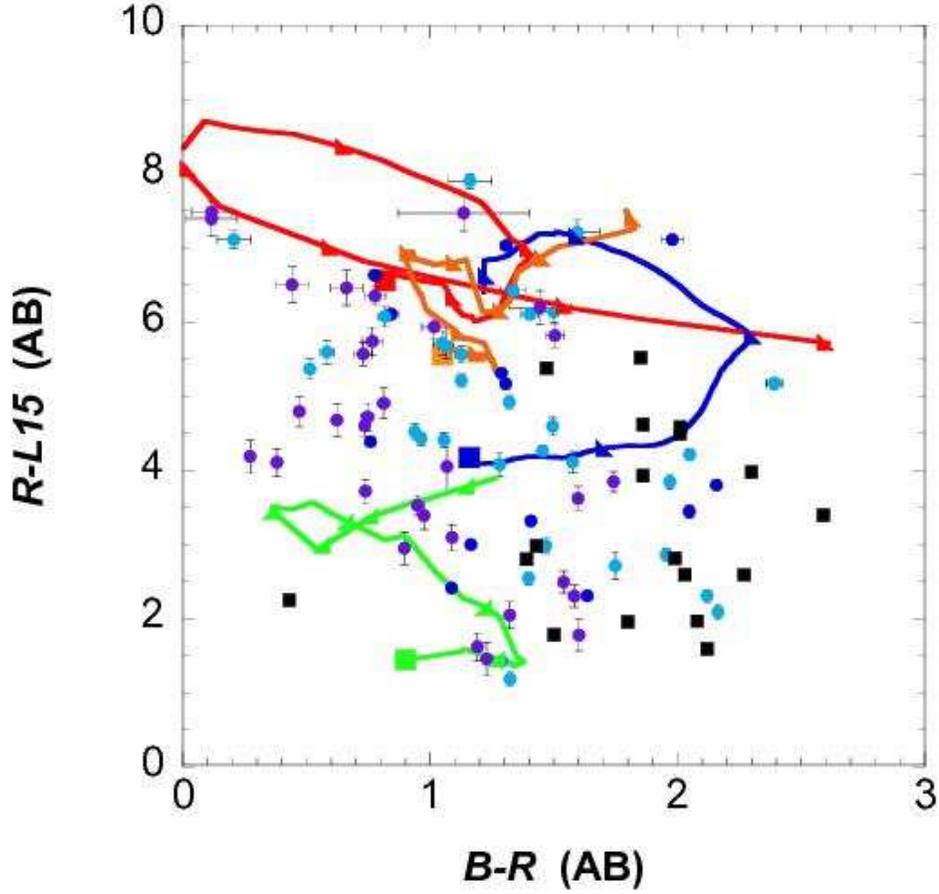}
  \end{center}
  \caption{
$B-R$ vs $R-L15$ plot for the 15~$\mu$m sources with successful identifications in the Subaru/Suprime-cam source catalog. Dark-blue, light-blue, \& purple, circles represent 16-18, 18-19, 19-20 AKARI 15$\mu$m AB magnitudes respectively. Color-color tracks for 4 SED templates are presented. Black squares represent the HDF-S sources detected with the ISOCAM LW3 deep survey~(\cite{oli02}, \cite{man02}). Green: a normal quiescent spiral galaxy (M51 template), blue: a star-forming galaxy (M82 template), red: an Ultraluminous infrared galaxy (ULIRG) Arp220 template, orange: an ULIRG, HR10 template. The large coloured squares are the zero redshift points for the SED templates and the markers along the template color tracks represents steps of 0.5 in redshift. 
 }\label{color-color}
\end{figure}

\begin{figure}[htbp]
  \begin{center}
    \FigureFile(126mm,126mm){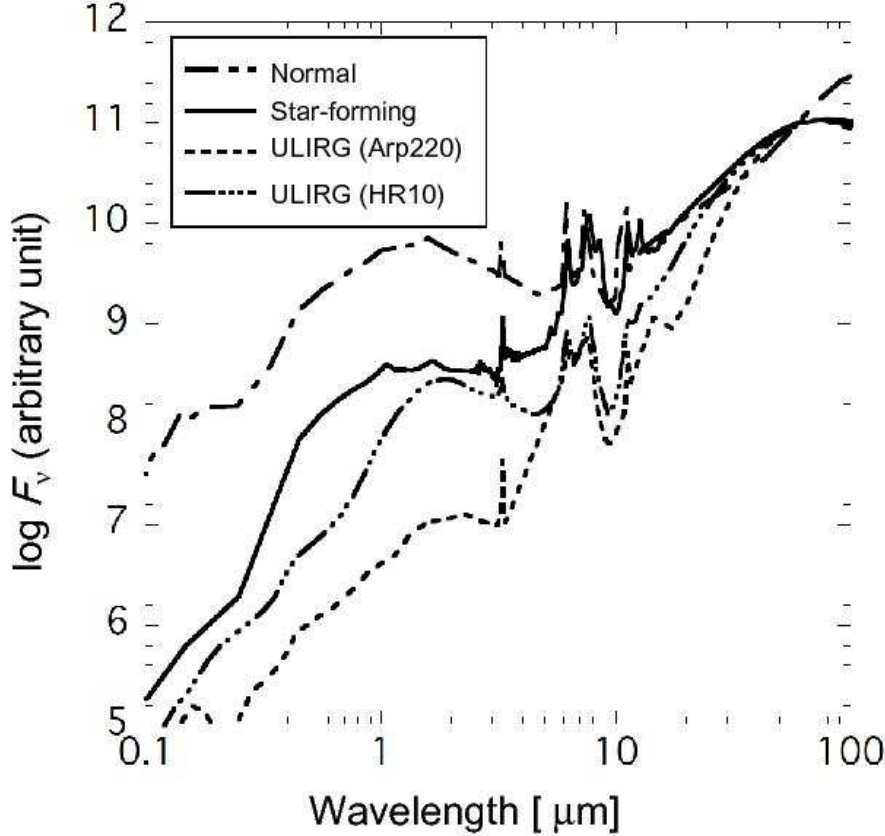}
  \end{center}
  \caption{
Rest-frame optical to far-infrared spectral energy distribution templates for; 
dashed line: a normal quiescent spiral galaxy (M51 template), 
solid line: a star-forming galaxy (M82 template), 
dotted line: an ultraluminous infrared galaxy (ULIRG), Arp220 template, 
dash-dash-dot-dot: an ULIRG, HR10 template. 
Model templates are taken from \citet{efstathiou03} for the normal galaxy template, 
\citet{efstathiou00} for the star-forming and Arp220 templates and \citet{takagi03b} for the HR10 template.
 }\label{seds}
\end{figure}

\begin{figure}[htbp]
  \begin{center}
    \FigureFile(126mm,126mm){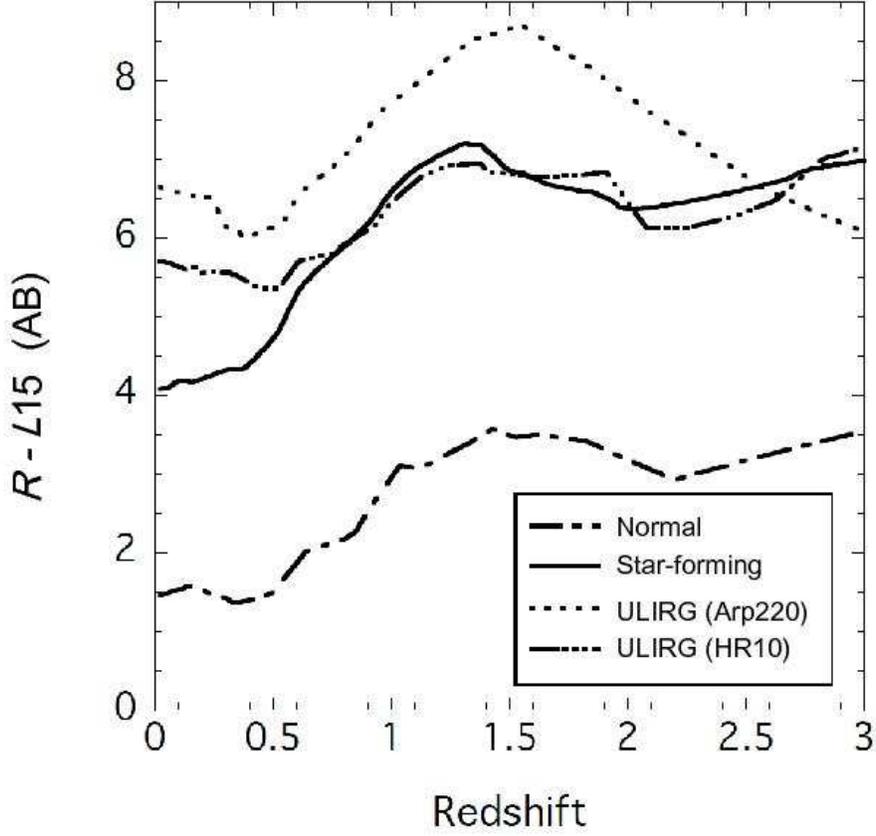}
  \end{center}
  \caption{
Optical to mid-infrared ($R-L15$ colour vs redshift for various template SEDs shown in Figure~\ref{seds}.
The SEDs are convolved with the filter-band response curves and then the colour is derived.
 }\label{color-z}
\end{figure}

In Figure \ref{color-color} we plot the $B-R$ vs $R-L15$ colours of the AKARI 15$\mu$m sources with successful identifications in 
the Subaru/Suprime-cam source catalog. The AKARI 15$\mu$m population are plotted over 3 magnitude ranges (15$\mu$m AB) from 16-18, 18-19 \&19-20th magnitude. The AKARI population spans a broad range in $R-L15$ colours from $R-L15$=1 to 8. To investigate the colours of the AKARI 15$\mu$m population we introduce a set of archetypal  galaxy spectral energy distribution (SED) templates. 
These templates are shown in Figure \ref{seds} and comprise a normal quiescent spiral galaxy modeled on the 
SED of M51, a dusty star-forming galaxy modeled on the SED of M82, and  ULIRGs modeled on the  SEDs of Arp220 
and HR10. The normal galaxy template is taken from the radiative transfer models of  \citet{efstathiou03} and 
the star-forming galaxy and ULIRG~(Arp220) templates are taken from the radiative transfer models of  
\citet{efstathiou00}. We also include another ULIRG template modeled on the SED of HR10 \citep{graham96} 
using the starburst spectral template library of \authorcite{takagi03a} (\yearcite{takagi03a}, 
\yearcite{takagi03b}). HR10 at $z$=1.44 is one of famous Extremely Red Objects~(EROs) and its best-fitting 
SED model is characterized by an older age and larger stellar mass than those of Arp220~\citep{takagi03b}. 
In Figure  \ref{color-z} we plot the $R-L15$ color of our model templates smoothed by the AKARI IRC filter--band response curves as a function of redshift. In general for any redshift, the dusty infrared sources are well separated from the normal quiescent galaxies, consistently having $R-L15 > $4. Figure  \ref{color-z} also suggests that sources with $R-L15 > $6 or 7 may be plausible candidates for redder dusty star-forming galaxies and ULIRGs respectively at 1$<$z$<$2. 
Since at $z$=1-2 the PAH emission features at 6.2 and 7.7~$\mu$m from the
star-forming galaxies enter the L15 passband and thus $R-L15$ colour becomes
redder, these red $R-L15$ colours may provide useful selection methods for
high redshift ULIRGs.
The $B-R$ vs $R-L15$ colours plotted in Figure  \ref{color-color} show that there is indeed 
a significant redder population of AKARI sources with $R-L15 >$4 that cannot be explained 
by the colours of normal quiescent galaxies or low-redshift star-forming galaxies. For these sources with 15$\mu$m AB magnitudes of 18, 19, this correspond to $R$ magnitudes of $>$22, $>$23 respectively.

 For the AKARI 15$\mu$m sources with successful identifications in 
the Subaru/Suprime-cam source catalog we find approximately 60 (40) per cent of the population with $R-L15 >$4 (5). There is also a significant fraction of the population (25 per cent) with very red colours of $R-L15 >$6. It is noteworthy that the sources detected in the ISOCAM deep surveys in the HDF-S~(\cite{oli02}, \cite{man02}) show mostly bluer ($R-L15 <$5) colours except for two sources, perhaps due to its shallower depth. These red sources populate the colour-colour parameter space occupied by the model SED templates for high redshift ULIRGs in Figure~\ref{color-color}.
 Note that \citet{mrr04} found that around $\sim$14 per cent of the sources  in the ELAIS final band merged catalogue ($S_{15\mu m}>$0.7mJy) could be attributed to ULIRGs. Given the deeper nature of the AKARI NEP survey, we may expect a significantly larger fraction of ULIRGs in our sample that most probably populate the  $R-L15 >$6 region of the  $B-R$ vs $R-L15$ colour-colour plane. This interpretation is also supported by the evolving mid-infrared luminosity function recently derived based on the Spitzer 24~$\mu$m deep surveys~(\cite{lef05}, \cite{pego05}).
Although we do not show the track for the AGN template SED, it should be noted that the power-law SEDs of dusty obscured AGN can satisfy $R-L15 >$5 for the power-law index 
$\alpha < -1.5$ ($f_{\nu} \propto \nu^{\alpha}$). \citet{aloh05} present a sample of such AGN SEDs from $\alpha=-0.5$ down to $\alpha=-2.8$. 
Thus, it is likely that a fraction of our 15$\mu$m sources may in fact be highly obscured AGN.

\begin{figure}[htbp]
  \begin{center}
    \FigureFile(126mm,126mm){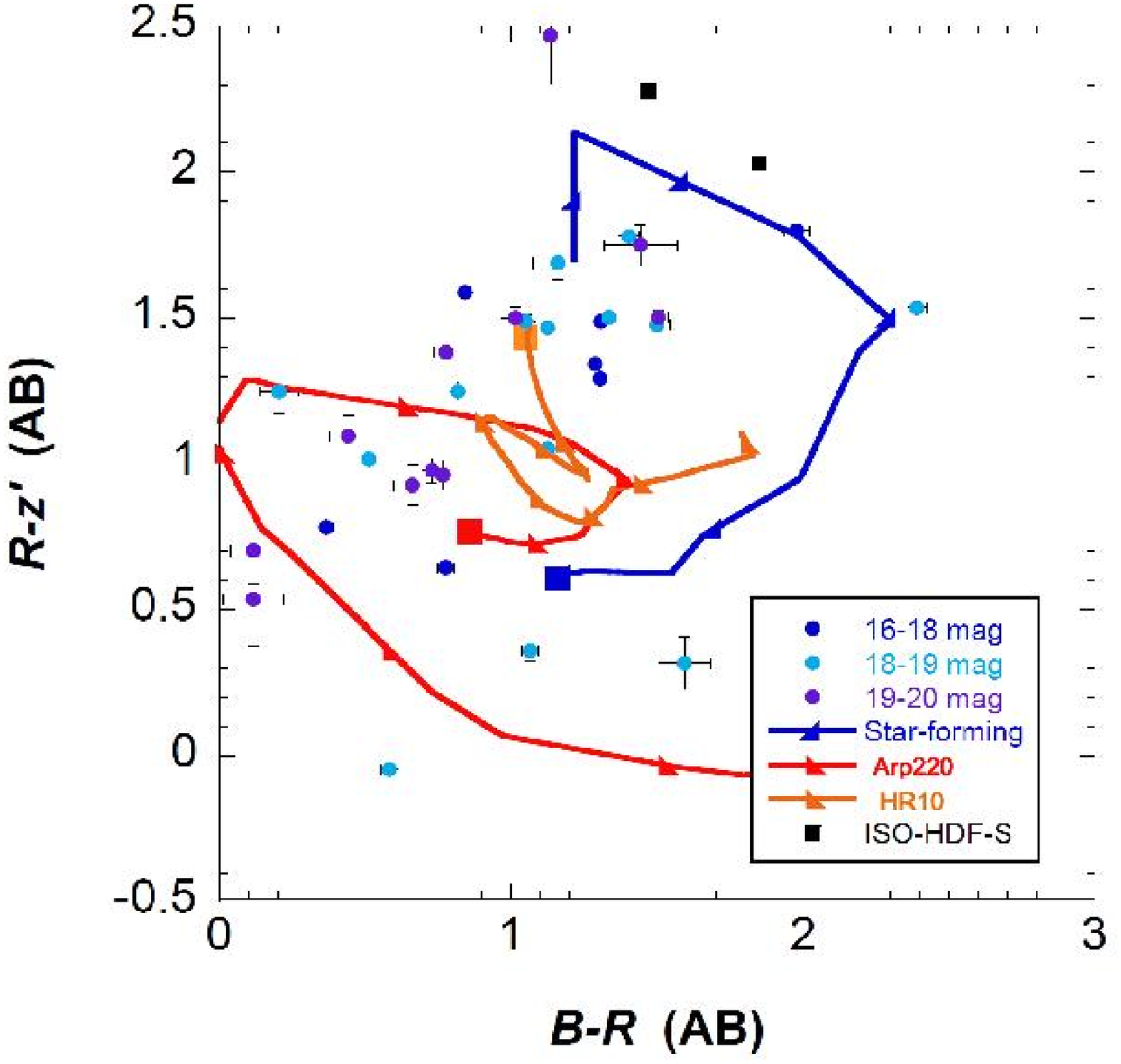}
  \end{center}
  \caption{
$B-R$ vs $R-z'$ plot for the 15~$\mu$m sources with $R-L15 > 5$ in the Subaru/Suprime-cam source catalog. Dark-blue, light-blue, \& purple, circles represent 16-18, 18-19, 19-20 AKARI 15$\mu$m AB magnitudes respectively. Two black squares represent the HDF-S sources~(\cite{oli02},\cite{man02}). Color-color tracks for 3 SED templates are presented: blue: a star-forming galaxy (M82 template), red: an ultraluminous infrared galaxy (ULIRG), Arp220 template, orange: an
ULIRG, HR10 template. The large coloured squares are the zero redshift points for the SED templates and the markers along the template color tracks represents steps of 0.5 in redshift. 
 }\label{color-color-optical}
\end{figure}

So far the spectroscopic redshifts for the red 15~$\mu$m sources are not available, and it is not easy to estimate their physical quantities, such as infrared luminosities, and stellar masses. However, for near-future follow-up studies, it is valuable to present a crude picture of the sources via optical colours. Figure~\ref{color-color-optical} shows a $B-R$ vs $R-z'$ plot for the 15~$\mu$m sources with $R-L15 > 5$, compared
with the SED templates of the starburst~(M82), ULIRGs~(Arp220, HR10). The red 15~$\mu$m sources are distributed widely over the colour parameter space, but at least we can say that the colours can be explained by the SED templates with $z \leq 2$. Of course the SED templates
presented here are not unique and there exists a wide variety of SED shapes for starburst galaxies (see also Takagi et al. in this volume). Thus optical and near infrared spectroscopic follow-ups with ground-based telescopes are essential to understand the
nature of these red 15~$\mu$m sources.

\subsection{Nature of the Selected 15~$\mu$m Sources}\label{sec:nature}

If the 15~$\mu$m sources are located at $z \geq 1$, quantitative studies of the 15~$\mu$m sources require near infrared ($J$ \& $K_{s}$-band) photometry, since the Balmer or 4000A break of the galaxy's SED is redshifted beyond optical wavelengths. For example, we may classify the 15~$\mu$m sources into either Extremely Red Objects~(\cite{manu02}, \cite{miya03})
or BzKs~\citep{dad04} based on their colour -- colour criteria. Unfortunately the $K_{s}$-band data available so far is found to be too shallow to extract the BzK population at
high redshift~($1.4 < z < 2.5$, \cite{dad04}). However we found six sources satisfying the ERO criterion ($ R - K_{s}>3$ in AB magnitude), and thus we attempt SED fitting of these sources in order to estimate their photometric redshifts, infrared luminosities, and SED types. 

We fitted the observed SEDs using the Bayesian photometric redshift code of \citet{beni00}.
 For the SED templates to compare with the data, we adopted 100  SEDs of 
 IR luminous galaxies from \citet{chael01}. These  SEDs are modified by applying
Galactic extinction corrections of $E(B-V)$ = 0.041 estimated from the far-infrared brightness of the NEP 
\citep{sch98} using the extinction curve of  \citet{calz00}.   
After fitting, we obtain an estimate of the photometric redshift as well as the 
best-fit SED template. The infrared luminosity is calculated by integrating the fluxes beyond the rest-frame 
5 $\mu$m of the best-fit SED template. 
We find that the formal errors in the photometric redshifts from the fitting program are under 20\%, and the uncertainty in the IR luminosity to be of order of at least a factor of a few. The estimates, however, might have a larger uncertainty if an object has a more complex SED than a simple, single-component SED as demonstrated for the object ID151 described below. We expect these uncertainties will be reduced when we have data between the $J$-band and the 15~$\mu$m, and also at 24~$\mu$m at the end of the NEP survey.

\subsubsection{Dusty Starburst Candidates}

\begin{figure}[htbp]
  \begin{center}
    \FigureFile(126mm,190mm){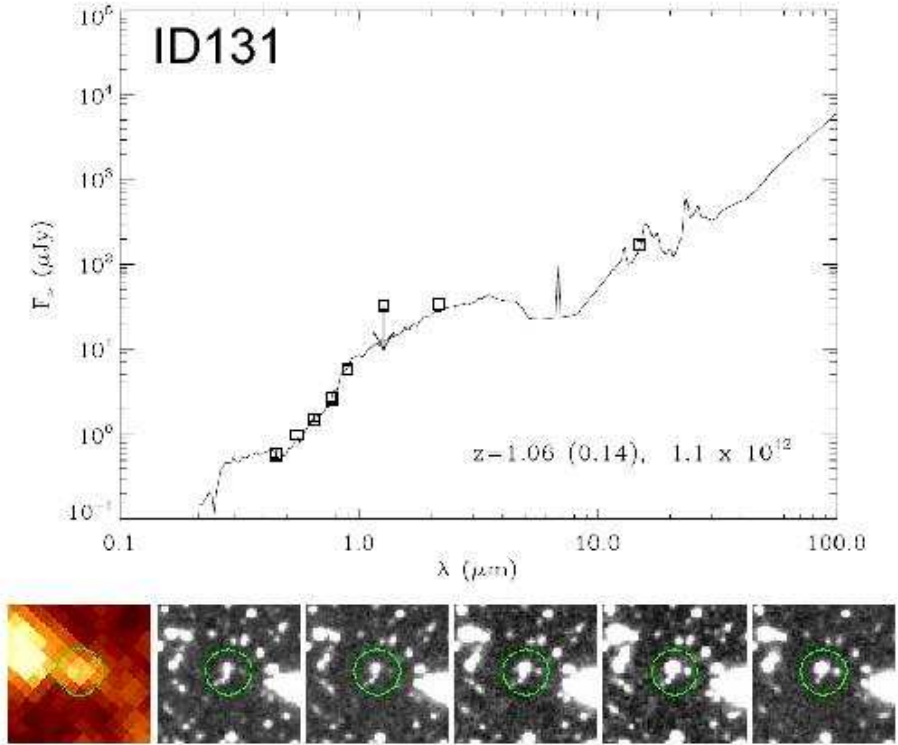}
  \end{center}
  \caption{
An example of the SED fitting (top) and the postage stamp images (bottom, $L15$, $B$, $V$, $R$, $i'$, and $z'$ from left to right) for sources with secure $K_{s}$ 
photometric data, in case of ID131. The radius of circles in all the postage stamps is 5~arcsec.
 }\label{sed_postage131}
\end{figure}

\begin{figure}[htbp]
  \begin{center}
    \FigureFile(126mm,190mm){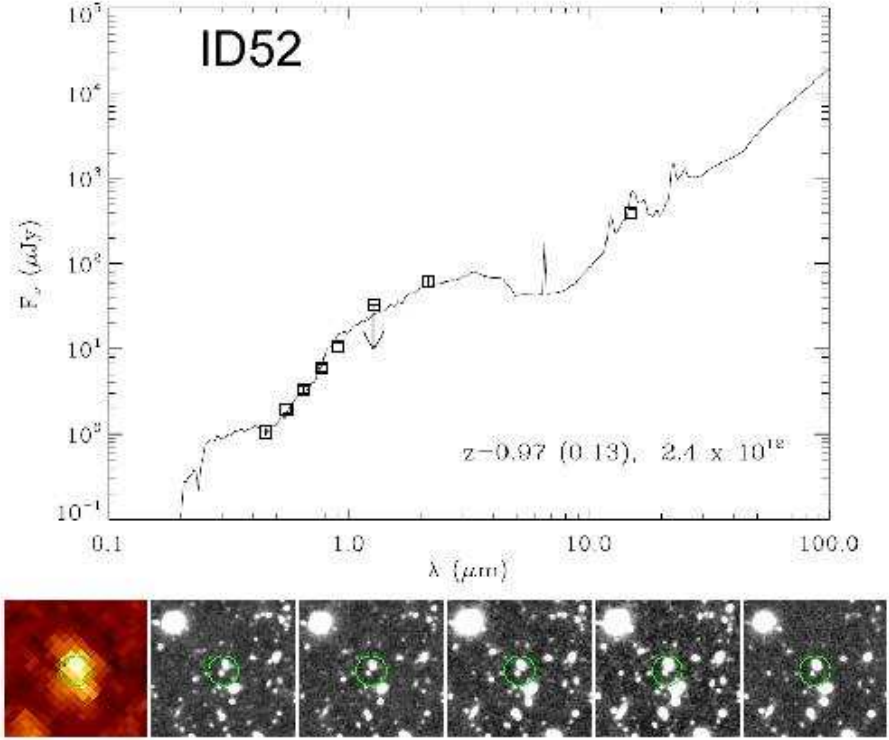}
  \end{center}
  \caption{
An example of the SED fitting (top) and the postage stamp images (bottom), in case of ID52. 
See Figure~\ref{sed_postage131} caption for explanation of figures.
 }\label{sed_postage52}
\end{figure}

Among the six 15~$\mu$m sources satisfying the ERO criterion, two sources (ID131 and ID52) can be fitted well with $z\sim1$ dusty starburst templates. In Figure~\ref{sed_postage131} and \ref{sed_postage52}, the results are shown: 
 for ID131,  the SED fitting gives a  photometric redshift~($z_{phot}$) of $1.06 \pm 0.14$, and total infrared luminosity~($L_{\rm IR}$) of $1.1\times10^{12}L_{\odot}$, while for ID52,  $z_{phot}=0.97 \pm 0.13$, $L_{\rm IR}=2.4\times10^{12}L_{\odot}$.

\subsubsection{E/S0 plus Dusty Starburst Composites?}

\begin{figure}[htbp]
  \begin{center}
    \FigureFile(126mm,190mm){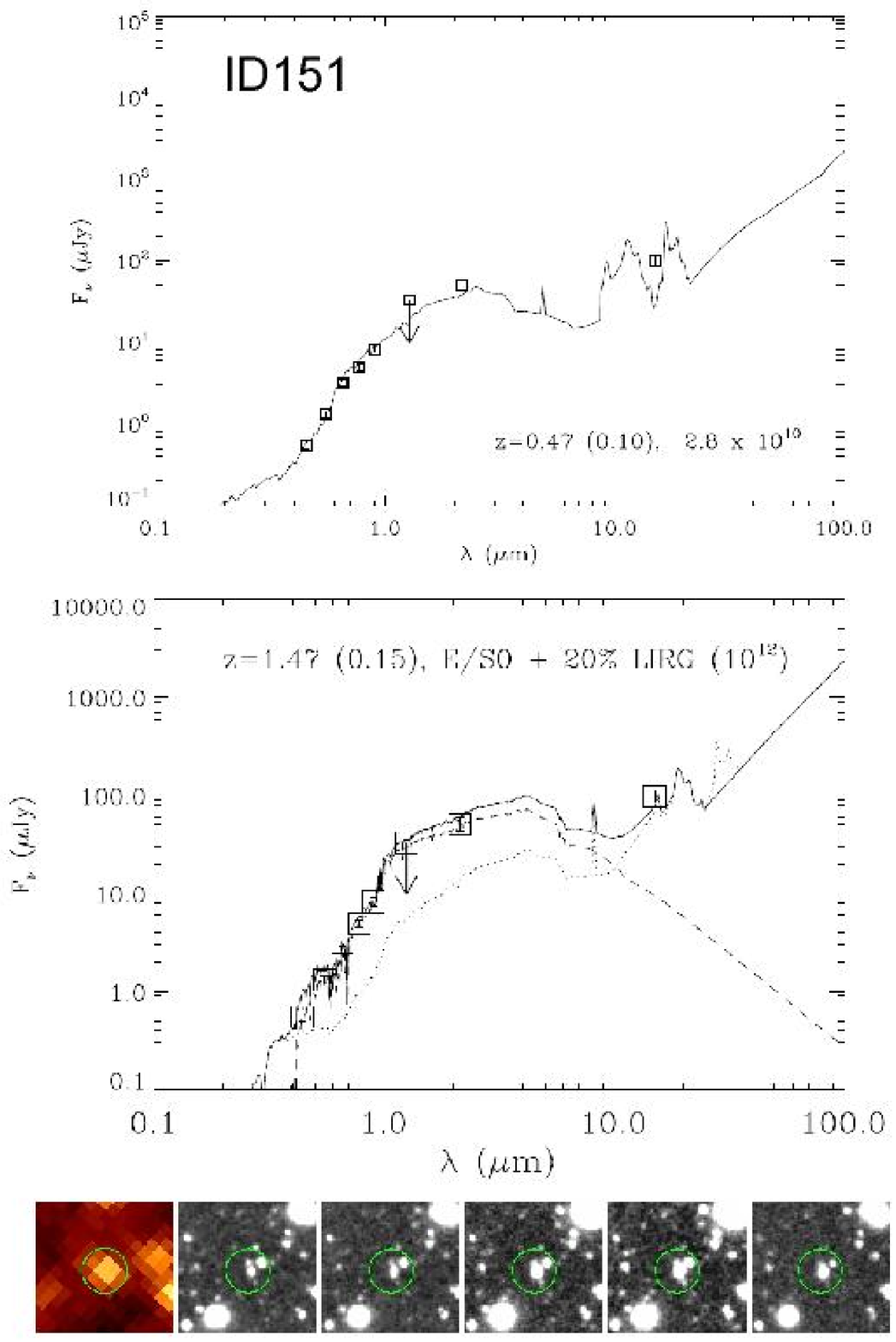}
  \end{center}
  \caption{
An example of the SED fitting (top two panels) and the postage stamp images (bottom) for the sources which can be fitted with not only by the dusty starburst templates but also by composites of E/S0 and dusty starburst SEDs. See Figure~\ref{sed_postage131} caption for explanation of figures.
 }\label{sed_postage151}
\end{figure}

The optical -- $K_{s}$-band SEDs of the final four 15~$\mu$m sources satisfying the ERO criterion can be fitted well by not only the dusty starburst templates at $z$=0.47~(Figure~\ref{sed_postage151} top) 
but also by old E/S0 like SEDs. 
However, for the latter case the 15~$\mu$m fluxes originating from the circumstellar dust of the AGB stars seems to be too weak to explain the observed fluxes. Therefore
 we propose that the SEDs can be explained by a composite SED of the E/S0-like stellar population plus the dusty starburst population. As shown in Figure~\ref{sed_postage151} middle, the SED of ID151 can be fitted by 0.8~Gyr-old (after a burst of star-formation over 0.1~Gyr produced by the 1996 version of \citet{brch93} models) stellar population and a dusty starburst with $L_{\rm IR}=10^{12}L_{\odot}$  which contributes 20\% of the $K_{s}$-band flux. In this case we obtain $z_{phot}= 1.47\pm 0.10$.
At the moment both fits seem to be acceptable. Measurements in the intermediate bands (3-11~$\mu$m) soon available from the AKARI ``NEP-Deep" survey will help to break this degeneracy. 

The AKARI NEP survey~\citep{maruma06} is continuing and will achieve  comparable mid-infrared depths to those described in this paper, over nine bands between 2 and 24~$\mu$m. Since the ``NEP-Deep" field covers approximately 20 times larger area than the performance verification field, we can expect to discover about 2000 faint ($\leq 100\,\mu$Jy) sources, from which we can construct statistically meaningful, optical -- mid infrared SED samples to further understand the nature of the 15~$\mu$m population introduced in this paper. 

\section{Summary}\label{sec:summary} 

The results of optical identifications are presented for 257 15~$\mu$m sources
 detected with a deep 15~$\mu$m survey over approximately 80~arcmin$^2$ area in
 the AKARI performance verification field around the North Ecliptic Pole. In comparison with the previous 15~$\mu$m surveys with ISO/ISOCAM and the
{\it Spitzer}/IRS peak-up imaging, the AKARI 15~$\mu$m sample is particularily unique in its
 faint flux limit ($\sim$40~$\mu$Jy) : the 15~$\mu$m fluxes of approximately a half of the sample are below 100~$\mu$Jy.
 Optical counterparts were searched for within a 2-3~arcsec search radius in both a $BVRi'z'$ catalog generated
from the deep Subaru/Suprime-cam field which covers one-third of
 the performance verification field, and the $g'r'i'z'$ catalog based on a wide-area survey made with MegaCam at CFHT.
We found that the $B-R$ and $R-z'$ colours of sources with successful optical identifications are systematically 
redder than that of the entire optical sample in the same field, indicating that the 15~$\mu$m sources may be located at relatively high redshift. Moreover, approximately 40\% of the 15~$\mu$m sources show colours $R-L15>5$, which cannot be explained by the SED of normal quiscent spiral
galaxies, but is consistent with the SEDs of redshifted ($z>1$) starburst or ULIRGs. 
This result indicates that the fraction of the ULIRGs in the faint 15~$\mu$m sample is much larger than that in the  brighter 15~$\mu$m sample. Based on optical to 15~$\mu$m SED fitting for a few sources with the $K_{s}$-band data available so far, we found that several 15~$\mu$m sources can be explained by an
SED of the dusty starburst population (ULIRGs).  Deep $J$ \& $K_{s}$-band data as well as AKARI mid-infrared multi-band data other than 15~$\mu$m are essential to further
constrain the nature of the faint 15~$\mu$m population. 

\section*{Acknowledgements} 
This work is based on observations with AKARI, a JAXA project with the
participation of ESA.
We would like to thank all AKARI team members for their support on this 
project. M.I. and E.K. were supported by the Korea Science and
  Engineering Foundation (KOSEF) grant funded by the Korea
  government (MOST), No. R01-2005-000-10610-0. 
H.M.L and M.G.L. were supported in part by ABRL (R14-2002-058-01000-0).
This work is partly supported by the JSPS grants (grant number 15204013, and 16204013).

\end{document}